%% file: template.tex
\documentclass{article}

\usepackage{arxiv}

\usepackage[utf8]{inputenc} 
\usepackage[T1]{fontenc}    
\usepackage{hyperref}       
\usepackage{url}            
\usepackage{booktabs}       
\usepackage{amsfonts}       
\usepackage{nicefrac}       
\usepackage{microtype}      
\usepackage{lipsum}

\usepackage{subcaption}
\usepackage[title,toc,titletoc,header]{appendix}
\usepackage{libertine}
\usepackage[libertine]{newtxmath}
\usepackage{listings}
\usepackage[ruled,vlined,linesnumbered]{algorithm2e}
\usepackage{algpseudocode}
\usepackage{array}
\usepackage{mdwmath}
\usepackage{mdwtab}
\usepackage{hyphenat}
\usepackage{balance}
\usepackage{comment}
\usepackage[font={small},skip=3pt]{caption}
  \DeclareCaptionType{copyrightbox}
\usepackage{graphicx}
\usepackage{hyperref}
\usepackage[capitalise,nameinlink,noabbrev]{cleveref}
\usepackage{listings}
\usepackage{paralist}
\usepackage{pgfplots}
\usepackage{tikz}
\usepackage{todonotes}
\usepackage{url}
\usepackage{verbatim}
\usepackage{xcolor}
\usepackage{environ}
\usepackage{wrapfig}
\usepackage{soul}
\usepackage{xspace}
\usepackage{multirow}
\usepackage{multicol}
\usepackage{xargs}
\usepackage{subcaption}
\usepackage{xr}
\usepackage{tabularx}
\usepackage{array}
\usepackage{soul}

\newcommand{\pluseq}{\mathrel{+}=}
\newcommand{\orcid}[1]{\href{https://orcid.org/#1}{\includegraphics[height=10pt]{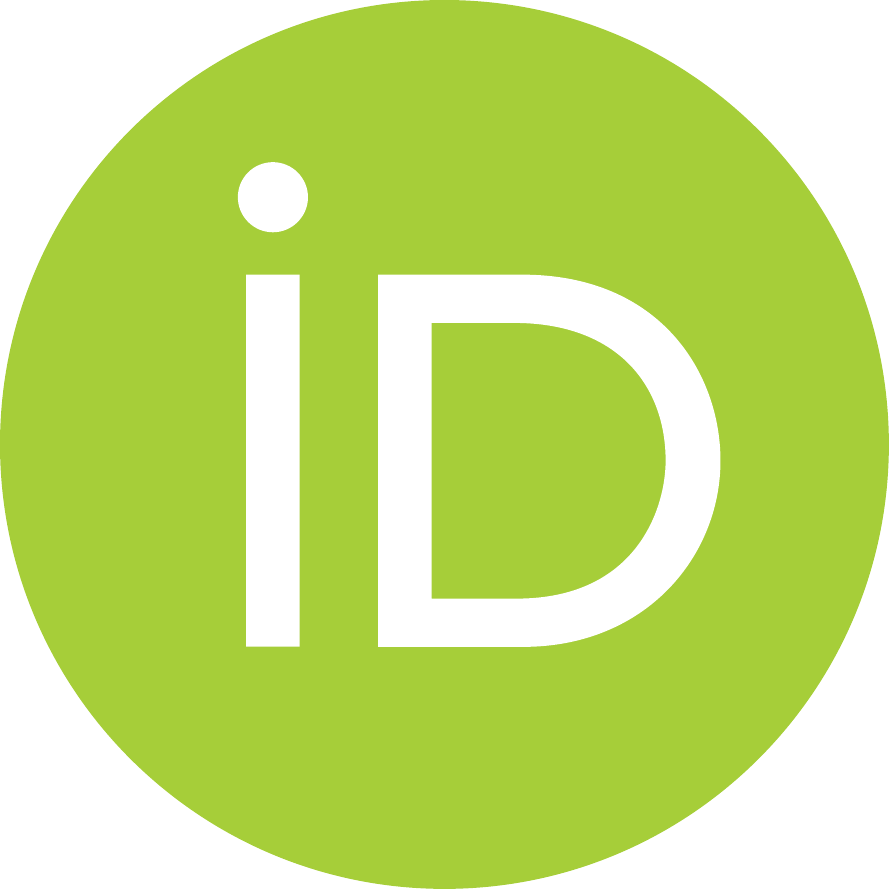}}}

\title{Memory Reduction using a Ring Abstraction over GPU RDMA for Distributed Quantum Monte Carlo Solver}

\author{
Weile Wei\orcid{0000-0002-3065-4959} \\
  Louisiana State University\\
  \texttt{wwei9@lsu.edu} \\
   \And
Eduardo D’Azevedo\orcid{0000-0002-6945-3206} \\
  Oak Ridge National Laboratory\\
  \texttt{dazevedoef@ornl.gov} \\
   \And
Kevin Huck\orcid{0000-0001-7064-8417} \\
  University of Oregon\\
  \texttt{khuck@cs.uoregon.edu} \\
\And
Arghya Chatterjee\orcid{0000-0002-7259-2944}\thanks{Arghya Chatterjee contributed to this work mostly during his previous appointment at Oak Ridge National Laboratory, Oak Ridge, TN.} \\
  NERSC, Berkeley Lab\\
  \texttt{ronnie@lbl.gov} \\
  \And
Oscar Hernandez\orcid{0000-0002-5380-6951} \\
  Oak Ridge National Laboratory\\
  \texttt{oscar@ornl.gov} \\
\And
Hartmut Kaiser\orcid{0000-0002-8712-2806} \\
  Louisiana State University\\
  \texttt{hkaiser@cct.lsu.edu} \\
}

\begin{document}
\maketitle

\input{sections/abstract}
\input{sections/introduction}
\input{sections/background}
\input{sections/implementation}
\input{sections/results}

\input{sections/discussion}
\input{sections/conclusion}
\input{sections/acknowledgement}

\bibliographystyle{unsrt}  

\bibliography{references}

\end{document}

%% file: sections/abstract.tex
\begin{abstract}
\label{sec:abstract}

Scientific applications that run on leadership computing facilities often face the challenge 
of being unable to fit leading science cases onto accelerator devices due to memory constraints 
(memory-bound applications).
%
In this work, the authors studied one such US Department of Energy mission-critical condensed matter 
physics application, Dynamical Cluster Approximation (DCA++), and this paper discusses how device memory-bound challenges were successfully reduced  by proposing an effective 
``all-to-all'' communication method---a ring communication algorithm. 
This implementation takes advantage of acceleration on GPUs and remote direct memory access (RDMA) for fast data exchange between GPUs. 
\\Additionally, the ring algorithm was optimized with sub-ring communicators
and multi-threaded support to further reduce communication overhead and 
expose more concurrency, respectively.
%
The computation and communication were also analyzed 
by using the Autonomic Performance Environment for Exascale 
(APEX) profiling tool,  and this paper further discusses the 
performance trade-off for the ring algorithm implementation. 
The memory analysis on the ring algorithm shows that the allocation size for the authors' most 
memory-intensive data structure per GPU is now reduced to $1/p$ of the original size, where $p$ is the number of GPUs in the ring communicator.
The communication analysis suggests that 
the distributed Quantum Monte Carlo execution time grows linearly as sub-ring size increases, and the cost of messages passing through the network interface connector could be a limiting factor.

%

\end{abstract}

\keywords{DCA++, Quantum Monte Carlo, GPU Remote Direct Memory Access, memory-bound issue, exascale machines}

%% file: sections/introduction.tex
\section{Introduction}
\label{sec:intro}

Dynamical Cluster Approximation (DCA++)\footnote{DCA++ is available at \url{https://github.com/CompFUSE/DCA}} is a high-performance research software 
application~\cite{PhysRevB.58.R7475, PhysRevB.61.12739, RevModPhys.77.1027, dca_hpx_2020} that provides a modern
C++ implementation to solve quantum many-body problems. 
DCA++ implements a quantum cluster method with a
Quantum Monte Carlo (QMC) kernel for modeling strongly
correlated electron systems. 
The DCA++ software currently uses three different
programming models---message passing interface (MPI), Compute Unified Device Architecture (CUDA), and High Performance ParalleX (HPX)/C++ threading---together with three numerical libraries---Basic Linear Algebra Subprograms (BLAS), Linear Algebra Package (LAPACK),
and Matrix Algebra on GPU (MAGMA)---to expose the parallel computation. 

In the QMC kernel~\cite{dca2019}, the two-particle Green's function ($G_t$) 
is needed for computing important fundamental quantities, such as
the critical temperature ($T_c$), for superconductivity. 
In other words, a larger $G_t$ allows condensed matter
physicists to explore larger and more complex (i.e., higher fidelity)
physics cases.
DCA++ currently stores $G_t$  in a single GPU device.
However, this limits the largest $G_t$ that can be processed within one
GPU.
A new approach for partitioning the large $G_t$ across 
the multiple GPUs can significantly increase 
scientists' capabilities to explore higher fidelity simulations.
This paper focuses on how the memory-bound issue in DCA++ was successfully addressed by proposing an effective ``all-to-all'' communication
method---a ring algorithm---to update the distributed $G_t$ device array.
\subsection{Contributions}
The primary contributions of this work are outlined as follows.
\begin{enumerate}
    \item The memory consumption in a QMC solver application was reduced to store a much larger $G_t$ array across multi-GPUs. This significant contribution 
    enables physicists to evaluate larger scientific problem sizes and 
    compute the full $G_t$ array in a single computation, 
    which significantly increases the accuracy/fidelity of the simulation of a certain material.
    \item A ring abstraction layer was designed that updates the
    large distributed $G_t$ array. The ring algorithm was further improved 
    by adding sub-ring communicator and multi-threaded communication to 
    reduce communication overhead and expose more concurrency, respectively.
    \item The ring abstraction layer was implemented on top of NVIDIA 
    GPUDirect remote direct memory
access (RDMA) for fast device memory transfer.
    \item The Autonomic Performance Environment for Exascale (APEX) performance measurement library was extended to support the use case, driving tool development and research.
\end{enumerate}

%% file: sections/background.tex
\section{Background}
\label{sec:background}

QMC solver applications are widely used and are mission-critical across the US Department
of Energy's (DOE's) application landscape. For the purpose of this paper, the authors chose to use one of the
primary QMC applications,
the DCA++ code.
A production-scale scientific
problem runs on DOE's fastest supercomputer, Summit, at the Oak Ridge Leadership Computing Facility on all 4,600
nodes; each node contains six NVIDIA Volta V100 GPUs, attaining a peak performance of 73.5 PFLOPS with a mixed
precision implementation~\cite{dca2019}.

\begin{figure}[h]
	\centering
	\includegraphics[width=\columnwidth]{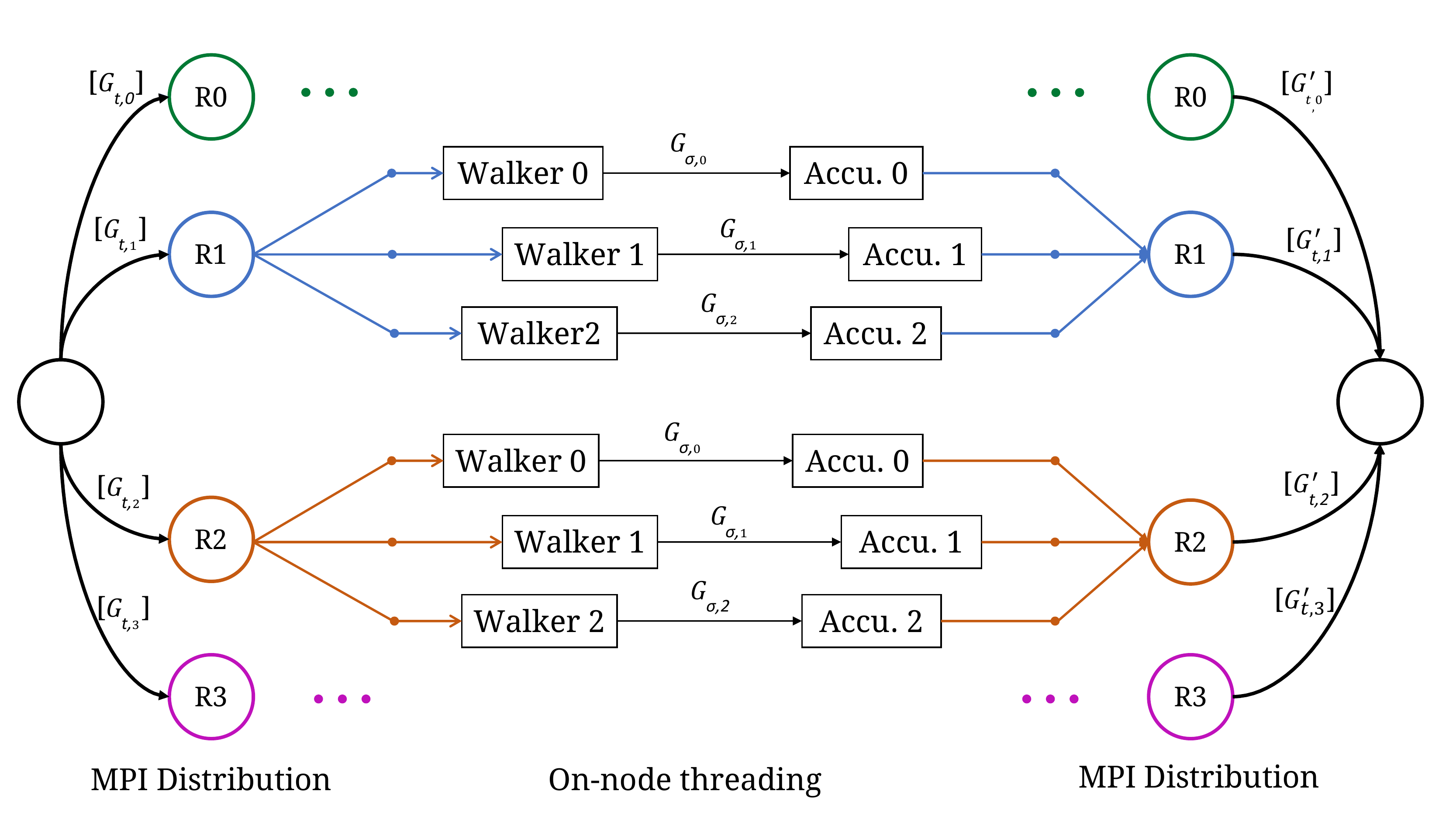}
	\caption{Workflow of the QMC DCA++ solver.}%
	\label{fig:dca_overview}
\end{figure}

Monte Carlo simulations are embarrassingly parallel, and the authors exploited this on distributed systems with a two-level
(MPI + threading) parallelization scheme (Figure~\ref{fig:dca_overview}). Although DCA++ has 
been highly optimized and is scalable on existing hardware, this is the first effort to focus on 
solving the memory-bound issue described in Section \ref{sec:intro} and further take advantage 
of Summit's GPU RDMA capability. 

Figure~\ref{fig:dca_overview} shows the parallelism
hierarchy in one iteration of the QMC solver (MPI
distribution + on-node threading parallelism). 
For example, each rank $\{R0,\ldots,RN\}$ 
is assigned a Markov Chain and the initial input (two particle
Green's function, $G_{t,i}$, where \textit{t} means ``two-particle,'' 
and \textit{i} is rank index).  Each rank spawns multiple independent
worker threads (i.e., walkers and accumulators). 
Most work and computation are performed on the GPU. 
Each walker thread generates a measurement
result ($G_{\sigma, i}$ array, where \textit{i} is thread ID) by performing 
nonuniform Fourier transform implemented by matrix-matrix multiplication.
Each walker passes its $G_{\sigma, i}$ to its corresponding accumulator
thread. 
In other words, each thread has its own $G_{\sigma, i}$ array,
and each rank will have $k$ different $G_{\sigma, i}$ arrays, where $k$
is the number of walker threads per rank. 
Each accumulator thread then updates $G_{t,i}$ via the formula in Eq.~(\ref{eq:G4})
to compute and update rank-local
$G_{t,i}$ to $G^{\prime}_{t,i}$.  
The updated partial $G^{\prime}_{t,i}$ is then fed
into the coarse-graining step for the next measurement. At the end
of all measurements, an \texttt{MPI\_Reduce} operation will be
performed on $G_t$ across all ranks to produce a final and complete
$G_{t}$ in the root rank. 
$G_t$ is allocated before all measurements start and has a life that spans
until the end of the DCA++ program.
%


\subsection{Memory-Bound Issue in DCA++}
The results from Balduzzi et al. \cite{dca2019} show that although storing a $G_t$ on the accelerator device allows condensed matter scientists
to explore larger and more complex (i.e., higher fidelity) physics
cases, the problem size is limited to the device memory size.
Updating the device array $G_t$ is the most time-consuming and memory-intensive process throughout DCA++ computation. A distributed $G_t$ approach is needed to reduce memory allocation and operation in the device.

In the original DCA++ algorithm, $G_t$ is 
updated by a product of two smaller
matrices (single-particle Green's function, or $G_{\sigma}$). This computation update is 
in the particle-particle channel and is accumulated according to Eq.~(\ref{eq:G4}).
\begin{equation}
\label{eq:G4}
    G_t (K_1, K_2, K_3)  \pluseq 
      \sum_{\sigma} G_\sigma (K_3 - K_2, K_3 - K_1) \, G_{-\sigma} (K_2, K_1)\,, 
\end{equation}
where $K_i$ is a combined index that represents a particular point in the momentum and frequency space, 
and $\sigma= {+1} \mbox{ or } {-1} $ specifies the electron spin value. $G_{\sigma}$ is the single-particle Green's function that
describes the configuration of single electrons.

%

The ability to handle a larger $G_t$  allows simulations of complex materials to significantly increase the details, accuracy, and fidelity.
In the previous design that kept $G_t$ within one GPU, only a
sub-slice of $G_t$ could be computed in a single computation.
For the simple single-orbital coarse-grained Hubbard model, physics insights or prior
knowledge can be used to decide which sub-slices in $G_t$
contain the important physics and thus avoid the generation of full $G_t$.
This simple model allows the generic behavior that comes
from electronic corrections in materials to be studied, but it cannot distinguish between
different specific materials.
Material-specific modeling requires more complex models
that include more orbital---and other---degrees of freedom, and
this requires a much larger $G_t$. 
This new distributed ring implementation enables the full large $G_t$
array to be computed in a single computation, even for more
complex multi-orbital models, to ensure that no important 
physics cases are overlooked.

\subsection{GPU RDMA Technology}
GPU RDMA allows direct peer access to multi-GPU memory through a high-speed network. 
For NVIDIA GPUs, GPUDirect is a technology that allows for the direct transfer of data in GPU device memory to other GPUs on the same node by using the NVLINK2 interconnect and/or between GPUs on different nodes by using RDMA support that can bypass buffers on host memory. 

A CUDA-aware MPI\footnote{\url{https://developer.nvidia.com/blog/introduction-cuda-aware-mpi/}} implementation can directly pass the GPU buffer pointer to MPI calls. Acceleration support, such as GPUDirect, can be used by the MPI library 
and allows buffers being sent from the kernel memory to a network without 
staging through host memory. There are various CUDA-aware MPI implementations, such as OpenMPI, MVAPICH2, and 
IBM Spectrum MPI\footnote{IBM Spectrum MPI is supported on
the Summit supercomputer, and is also the CUDA-aware MPI implementation used by the authors in this paper.}.

%% file: sections/implementation.tex
\section{Implementation: Ring Abstraction}
\label{sec:implement}
\subsection{Distributed \mbox{$G_t$} in QMC Solver}
\label{distributedG4}
\begin{figure}
\centering
     \begin{subfigure}[b]{\columnwidth}
         \centering
         \includegraphics[width=\textwidth]{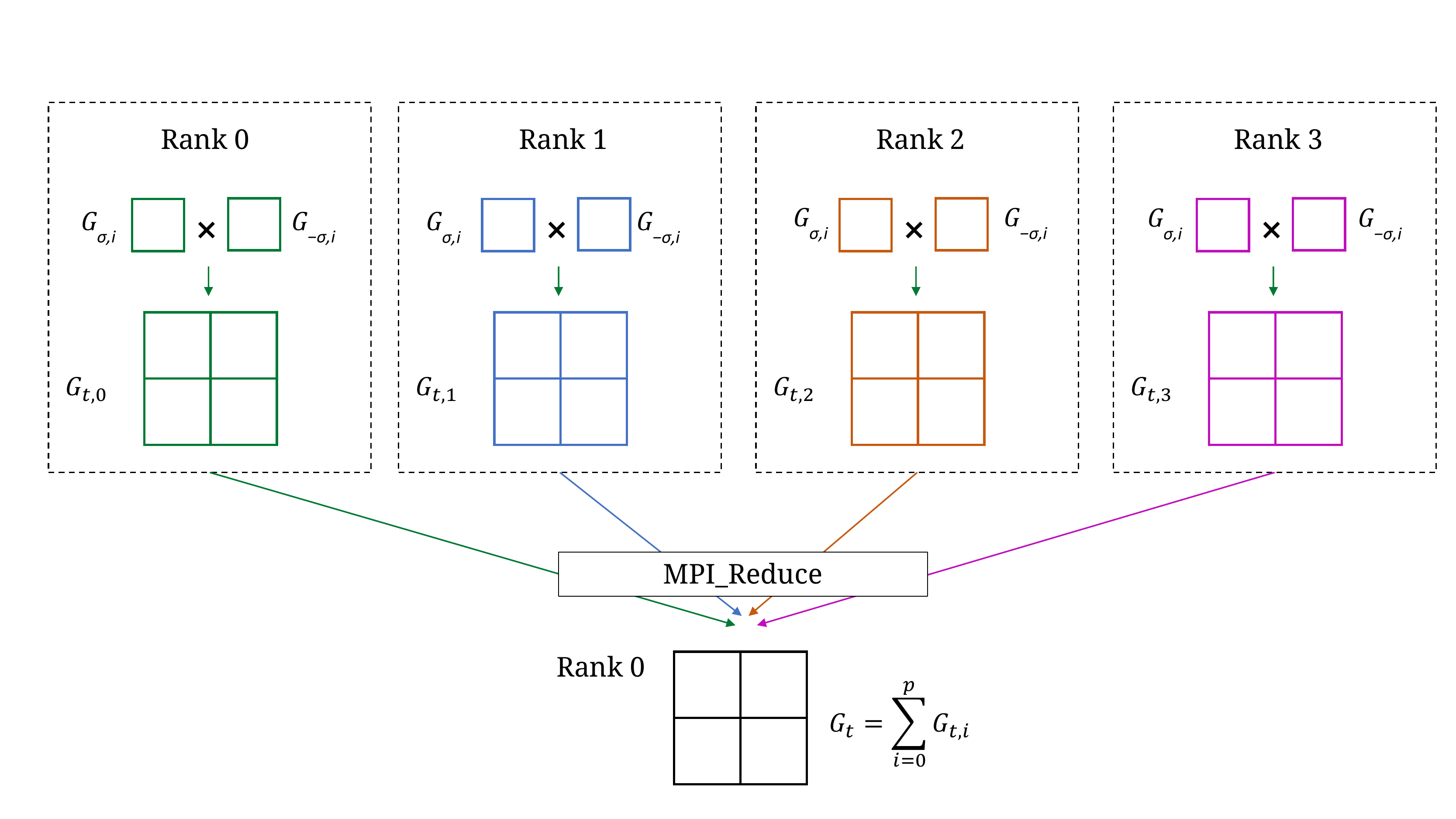}
         \caption{Original $G_t$ implementation.}
         \label{fig:original_g4}
     \end{subfigure}
     
    \begin{subfigure}[b]{\columnwidth}
         \centering
         \includegraphics[width=\textwidth]{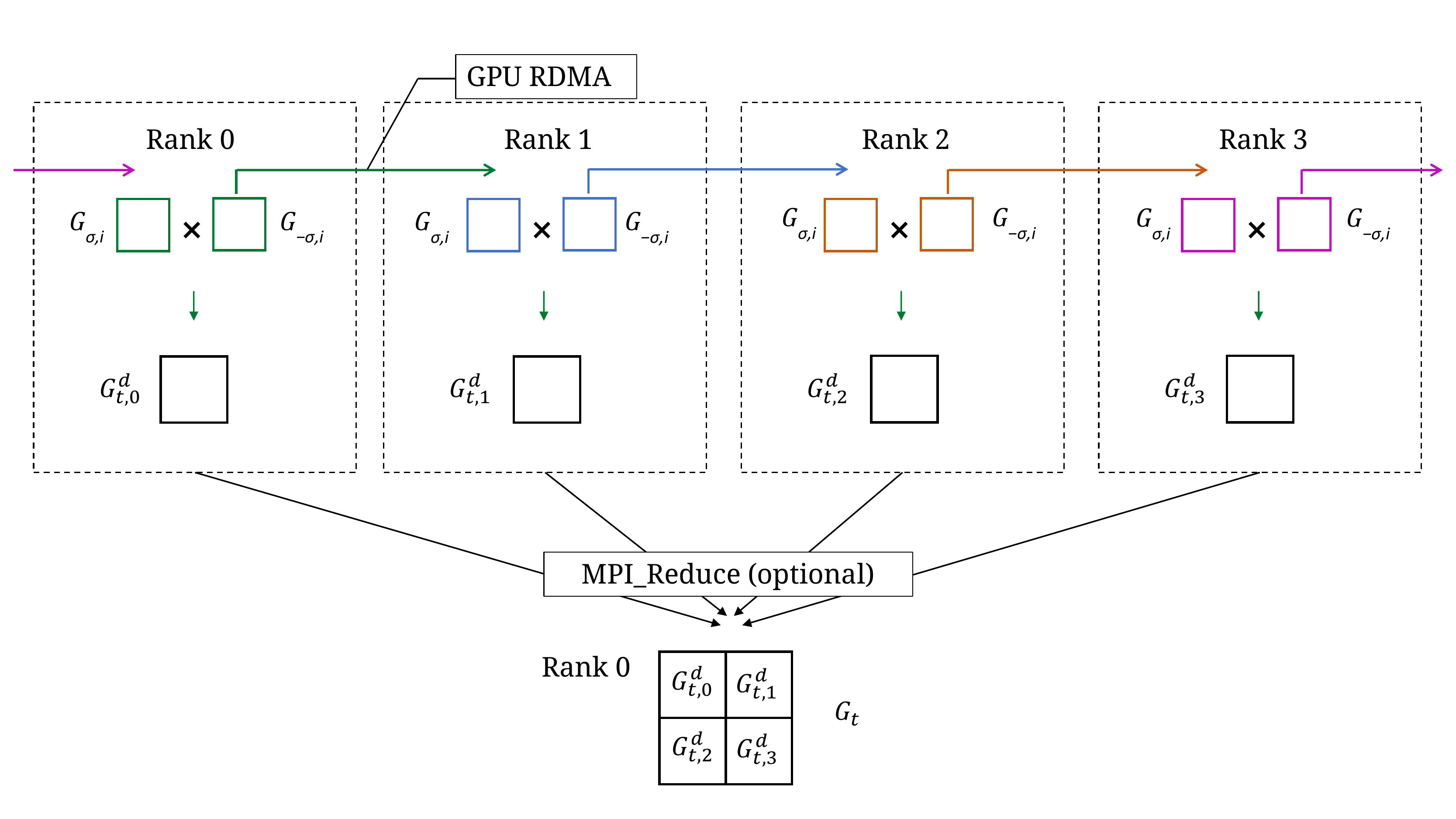}
         \caption{Distributed $G_t$ implementation.}
         \label{fig:distributed_g4}
     \end{subfigure}
     
\caption{Comparison of the original $G_t$ vs. the distributed $G^d_t$ implementation. Each rank contains one GPU resource.}
\label{fig:compare_original_distributed_g4}
\end{figure}

Before introducing the communication phase of the ring abstraction layer,
it is important to understand how the authors distributed the large device array $G_t$ across MPI ranks.
Original $G_t$ was compared, and $G^d_t$ versions were distributed
(Figure~\ref{fig:compare_original_distributed_g4}).

In the original $G_t$ implementation, the measurements---which were computed by matrix-matrix multiplication---are distributed statically and independently over the MPI ranks to avoid
inter-node communications. Each MPI rank keeps its partial copy of $G_{t,i}$ to accumulate 
measurements within a rank, where $i$ is the rank index. 
After all the measurements are finished, a reduction step is 
taken to accumulate $G_{t,i}$ across all MPI ranks into a final and complete
$G_t$ in the root MPI rank. The size of the $G_{t,i}$ in each rank is 
the same size as the final and complete $G_t$. 

With the distributed $G^d_t$ implementation, this large device array 
$G_t$ was evenly partitioned across all MPI ranks; each portion of it is local to each MPI rank.
Instead of keeping its partial copy of $G_t$, 
each rank now keeps an instance of $G^d_{t,i}$ to accumulate measurements 
of a portion or sub-slice of the final and complete $G_t$, where the notation
$d$ in $G^d_t$  refers to the distributed version, and $i$ means the $i$-th rank.
The $G^d_{t,i}$ size in each rank is 
reduced to $1/p$ of the size of the final and complete $G_t$, comparing the same configuration 
in original $G_t$ implementation, where $p$ is the number of MPI ranks used. 
For example, in Figure~\ref{fig:distributed_g4}, there are four ranks, and rank $i$
now only keeps $G^d_{t,i}$, which is one-fourth the size of the original $G_t$ array size.

To compute the final and complete $G^d_{t,i}$ for the distributed $G^d_t$ implementation, 
each rank must see every $G_{\sigma,i}$ from all ranks. 
In other words, each rank must broadcast the
locally generated $G_{\sigma,i}$ to the remainder of the other ranks at every measurement step. 
To efficiently perform this ``all-to-all'' broadcast, a ring abstraction layer was built (Section. \ref{section:ring_algorithm}), which circulates
all $G_{\sigma,i}$ across all ranks.





\subsection{Pipeline Ring Algorithm}
\label{section:ring_algorithm}
\begin{figure}
	\centering
	\includegraphics[width=\columnwidth, trim=0 5cm 0 0, clip]{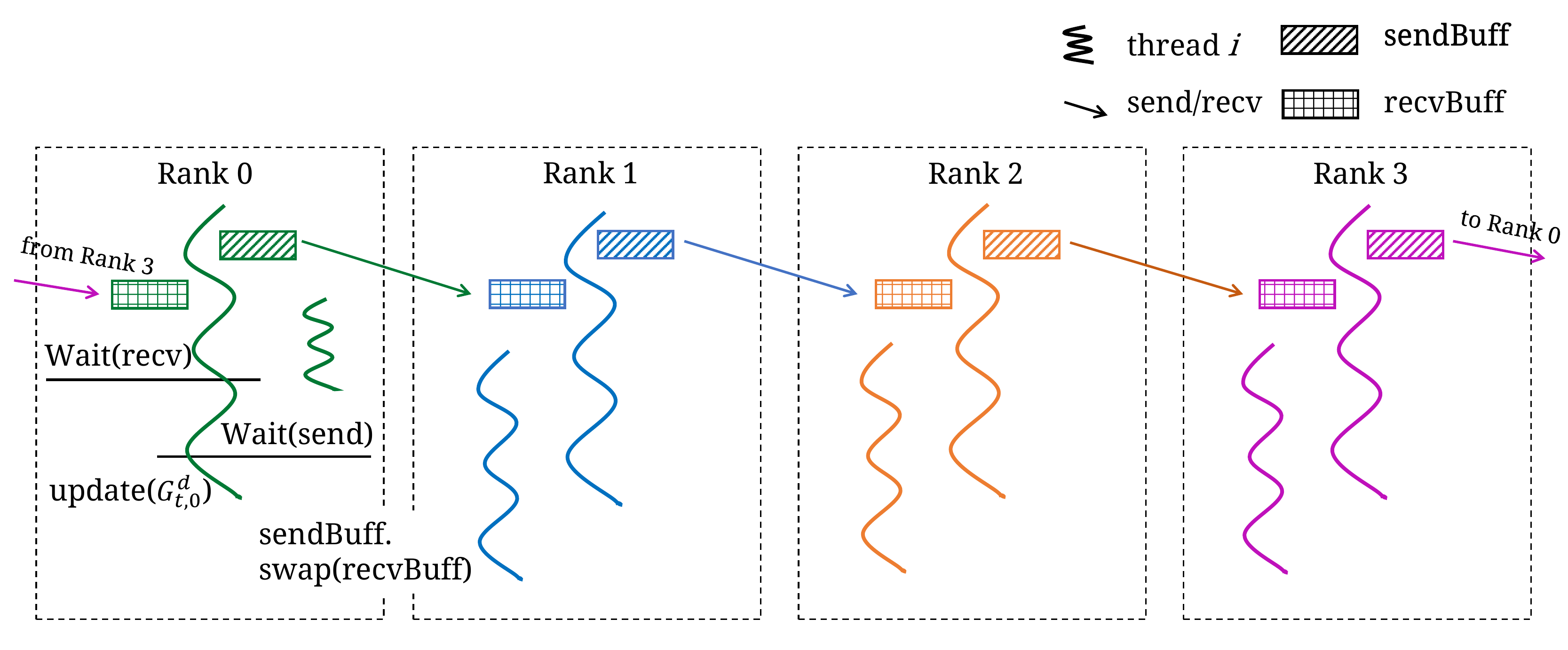}
	\caption{Workflow of ring algorithm per iteration. }
	\label{fig:ring_algorithm_figure}
\end{figure}

A pipeline ring algorithm was implemented that broadcasts the $G_{\sigma}$ 
array circularly during every measurement. 
The algorithm (Algorithm \ref{alg:ring_algorithm_code}) is 
visualized in Figure~\ref{fig:ring_algorithm_figure}.

\begin{algorithm}
\SetAlgoLined
    generateGSigma(gSigmaBuf)\; \label{lst:line:generateG2}
    updateG4(gSigmaBuf)\;       \label{lst:line:updateG4}
    {$i\leftarrow 0$}\;         \label{lst:line:initStart}
    {$myRank \leftarrow worldRank$}\;          \label{lst:line:initRankId}
    {$ringSize \leftarrow mpiWorldSize$}\;      \label{lst:line:initRingSize}
    {$leftRank \leftarrow (myRank - 1 + ringSize) \: \% \: ringSize $}\;
    {$rightRank \leftarrow (myRank + 1 + ringSize) \: \% \: ringSize $}\;
    sendBuf.swap(gSigmaBuf)\;           \label{lst:line:initEnd}
    \While{$i< ringSize$}{
        MPI\_Irecv(recvBuf, source=leftRank, tag = recvTag, recvRequest)\; \label{lst:line:Irecv}
        MPI\_Isend(sendBuf, source=rightRank, tag = sendTag, sendRequest)\; \label{lst:line:Isend}
        MPI\_Wait(recvRequest)\;        \label{lst:line:recevBuffWait}
        
        updateG4(recvBuf)\;             \label{lst:line:updateG4_loop}
        
        MPI\_Wait(sendRequest)\;        \label{lst:line:sendBuffWait}
        
        sendBuf.swap(recvBuf)\;         \label{lst:line:swapBuff}
        i++\;
    }
\caption{Pipeline ring algorithm}
\label{alg:ring_algorithm_code}
\end{algorithm}

At the start of every new measurement, a single-particle Green's function $G_{\sigma}$
 (Line~\ref{lst:line:generateG2}) is generated 
and then used to update $G^d_{t,i}$ (Line~\ref{lst:line:updateG4})
via the formula in Eq.~(\ref{eq:G4}).
%
%
Between Lines \ref{lst:line:initStart} to \ref{lst:line:initEnd}, the algorithm 
initializes the indices
of left and right neighbors and prepares the sending message buffer from the
previously generated $G_{\sigma}$ buffer. 
The processes are organized as a ring so that the first and last rank are considered to be neighbors to each other. 
A \textit{swap} operation is used to avoid unnecessary memory copies for \textit{sendBuf} preparation.
A walker-accumulator thread allocates an additional \textit{recvBuf} buffer of the same size 
as \textit{gSigmaBuf} to hold incoming 
\textit{gSigmaBuf} buffer from \textit{leftRank}. 

The \textit{while} loop is the core part of the pipeline ring algorithm. 
For every iteration, each thread in a rank 
receives a $G_{\sigma}$ buffer from its left neighbor rank and sends a $G_{\sigma}$ buffer to its right neighbor rank. 
A synchronization step (Line~\ref{lst:line:recevBuffWait}) is performed
afterward to ensure that each rank receives a new buffer to update the 
local $G^d_{t,i}$ (Line~\ref{lst:line:updateG4_loop}). 
Another synchronization step
follows to ensure that all send requests are finalized 
(Line~\ref{lst:line:sendBuffWait}). Lastly, another \textit{swap} operation is used to exchange
content pointers between \textit{sendBuf} and \textit{recvBuf} to avoid unnecessary memory copy and prepare
for the next iteration of communication.
In the multi-threaded version (Section~\ref{subsec:multi-thread}), the thread of index, \textit{i}, only communicates with
	threads of index, \textit{i}, in neighbor ranks, and each thread allocates two buffers: \textit{sendBuff} and \textit{recvBuff}.

The \textit{while} loop will be terminated after $\mbox{\textit{ringSize}} - 1$ steps. By that time, 
each locally generated $G_{\sigma,i}$ will have traveled across all MPI ranks and
updated $G^d_{t,i}$ in all ranks. Eventually, each $G_{\sigma,i}$ reaches
to the left neighbor of its birth rank. For example, $G_{\sigma,0}$ generated from rank $0$ will end 
in last rank in the ring communicator.

Additionally, if the $G_t$ is too large to be stored in one node, 
it is optional to accumulate all $G^d_{t,i}$
at the end of all measurements. 
Instead, a parallel write into the file system could be taken.

\subsubsection{Sub-Ring Optimization.}

\begin{figure}
	\centering
	\includegraphics[width=\columnwidth, trim=0 5cm 0 0, clip]{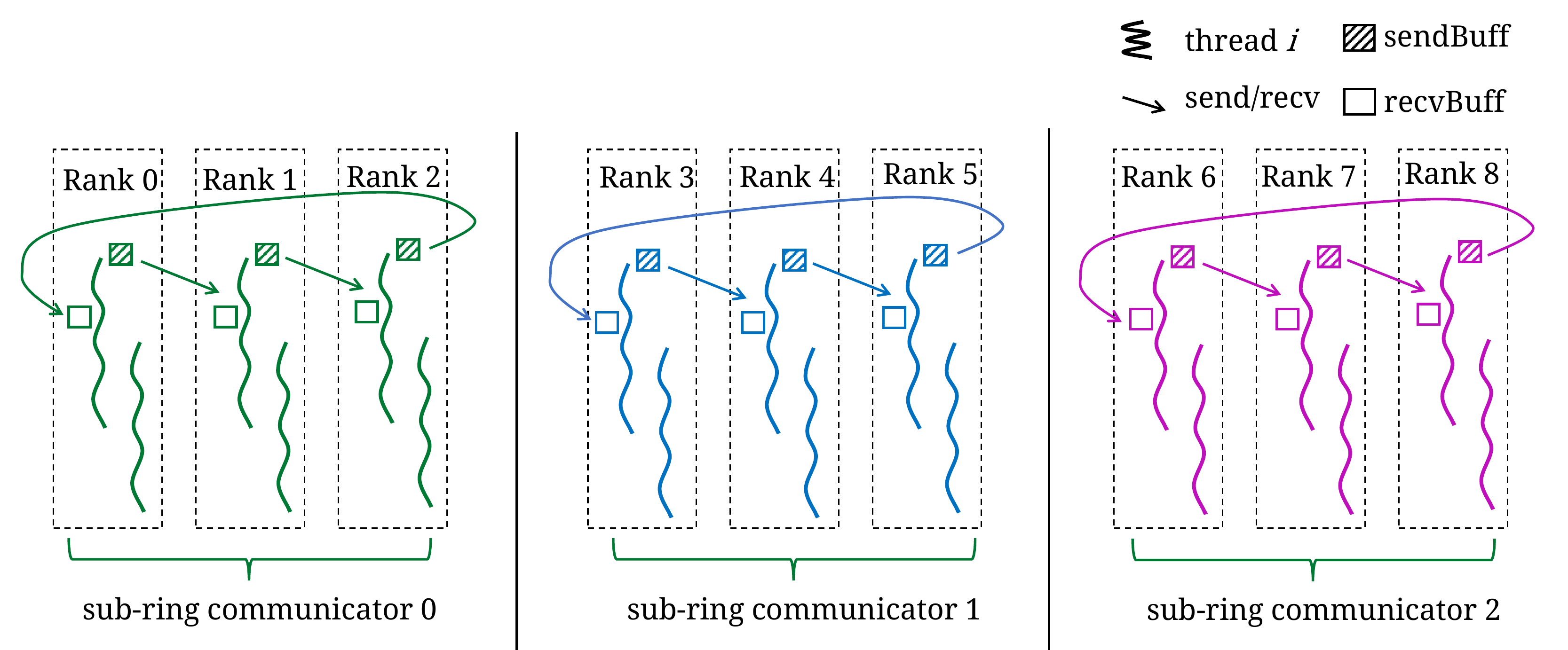}
	\caption{Workflow of sub-ring algorithm per iteration. Every consecutive $S$ rank forms a sub-ring communicator, 
	and no communication occurs between sub-ring communicators until all measurements are finished. Here, $S$ is the number of ranks in a sub-ring.}
	\label{fig:subring_algorithm_figure}
\end{figure}

A sub-ring optimization strategy is further proposed to reduce message communication
times if the large device array $G_t$ can fit in fewer devices. 
The sub-ring algorithm is visualized in Figure~\ref{fig:subring_algorithm_figure}.

For the ring algorithm (Section~\ref{section:ring_algorithm}), the size of the ring communicator
(\textit{mpiWorldSize}) is set to the same size of the global \mbox{\texttt{MPI\_COMM\_WORLD}}, and thus the size of $G_t$ is equally 
distributed across all MPI ranks.

However, to complete the update to $G^d_{t,i}$ in one measurement, 
one $G_{\sigma,i}$
must travel \textit{mpiWorldSize} ranks. In total, 
there are \textit{mpiWorldSize} numbers of $G_{\sigma,i}$
being sent and received concurrently in one measurement 
in the global
\mbox{\texttt{MPI\_COMM\_WORLD}} 
communicator. If the size of $G^d_{t,i}$ is relatively small per rank, then this will cause high communication overhead.

If $G_t$ can be distributed and fitted in fewer devices, then a shorter travel distance is required 
for $G_{\sigma,i}$, thus reducing the communication overhead. One reduction
step was performed at the end of all measurements to accumulate $G^d_{t,s_i}$, 
where $s_i$ means $i$-th rank on the $s$-th sub-ring.

At the beginning of MPI initialization, the global \mbox{\texttt{MPI\_COMM\_WORLD}} was partitioned  into several new sub-ring communicators by using \mbox{\texttt{MPI\_Comm\_split}}. 
The new
communicator information was passed to the DCA++ concurrency class by substituting the original global 
\mbox{\texttt{MPI\_COMM\_WORLD}} with this new communicator. Now, only a few minor modifications
are needed to transform the ring algorithm (Algorithm~\ref{alg:ring_algorithm_code})
to sub-ring Algorithm~\ref{alg:sub_ring_algorithm}. In Line~\ref{lst:line:initRankId}, \textit{myRank} is 
initialized to \textit{subRingRank} instead of \textit{worldRank}, where 
\textit{subRingRank} is the rank index in the local sub-ring communicator. 
In Line~\ref{lst:line:initRingSize}, \textit{ringSize} is initialized to \textit{subRingSize}
instead of \textit{mpiWorldSize}, where \textit{subRingSize} is the
size of the new communicator.
The general ring algorithm is a special case for the sub-ring algorithm because the
\textit{subRingSize} of the general ring algorithm is equal to \textit{mpiWorldSize}, and
there is only one sub-ring group throughout all MPI ranks.

\LinesNumberedHidden
\begin{algorithm}
    {$\mbox{\textit{myRank}} \leftarrow \mbox{\textit{subRingRank}}$}\;         
    {$\mbox{\textit{ringSize}} \leftarrow \mbox{\textit{subRingSize}}$}\;      
\caption{Modified ring algorithm to support sub-ring communication}
\label{alg:sub_ring_algorithm}
\end{algorithm}

\subsubsection{Multi-Threaded Ring Communication.}
\label{subsec:multi-thread}
To take advantage of the multi-threaded QMC model already in DCA++, 
multi-threaded ring communication support was further implemented in the ring algorithm.
Figure~\ref{fig:dca_overview} shows that in the original DCA++ method,
each walker-accumulator
thread in a rank is independent of each other, and all the threads in a 
rank synchronize only after all rank-local measurements are finished.
Moreover, during every measurement, each walker-accumulator thread
generates its own thread-private $G_{\sigma, i}$ to update $G_t$. 

The multi-threaded ring algorithm now allows concurrent message exchange so that threads of same rank-local thread index exchange their thread-private $G_{\sigma, i}$. 
Conceptually, there are $k$ parallel and independent rings, where $k$ 
is number of threads per rank, because threads of the same local thread ID
form a closed ring. 
For example, a thread of index $0$ in rank $0$ will send its $G_\sigma$ to 
the thread of index $0$ in rank $1$ and receive another $G_\sigma$ from thread index of $0$ 
from last rank in the ring algorithm.

The only changes in the ring algorithm are offsetting the tag values 
(\texttt{recvTag} and \texttt{sendTag}) by the thread index value. For example,
Lines~\ref{lst:line:Irecv} and ~\ref{lst:line:Isend} from 
Algorithm~\ref{alg:ring_algorithm_code} are modified to Algorithm~\ref{alg:multi_threaded_ring}.

\LinesNumberedHidden
\begin{algorithm}
        MPI\_Irecv(recvBuf, source=leftRank, tag = recvTag + threadId, recvRequest)\; 
        MPI\_Isend(sendBuf, source=rightRank, tag = sendTag + threadId, sendRequest)\;
\caption{Modified ring algorithm to support multi-threaded ring}
\label{alg:multi_threaded_ring}
\end{algorithm}

To efficiently send and receive $G_\sigma$, each thread
will allocate one additional \textit{recvBuff} to hold incoming 
\textit{gSigmaBuf} buffer from \textit{leftRank} and perform send/receive efficiently.
In the original DCA++ method, there are $k$ numbers of buffers of $G_\sigma$ 
size per rank, and in the multi-threaded ring method, there are $2k$
numbers of buffers of $G_\sigma$ size per rank, where $k$ is number of 
threads per rank.

%% file: sections/results.tex
\section{Results}
\label{sec:results}
This section evaluates this work from various perspectives---including correctness, memory analysis, scaling, and function activities---with help from the APEX profiling tool. All experiments were run on 
Summit.

\subsection{Summit Node Configuration}
\begin{figure}
	\centering
	\includegraphics[width=\columnwidth]{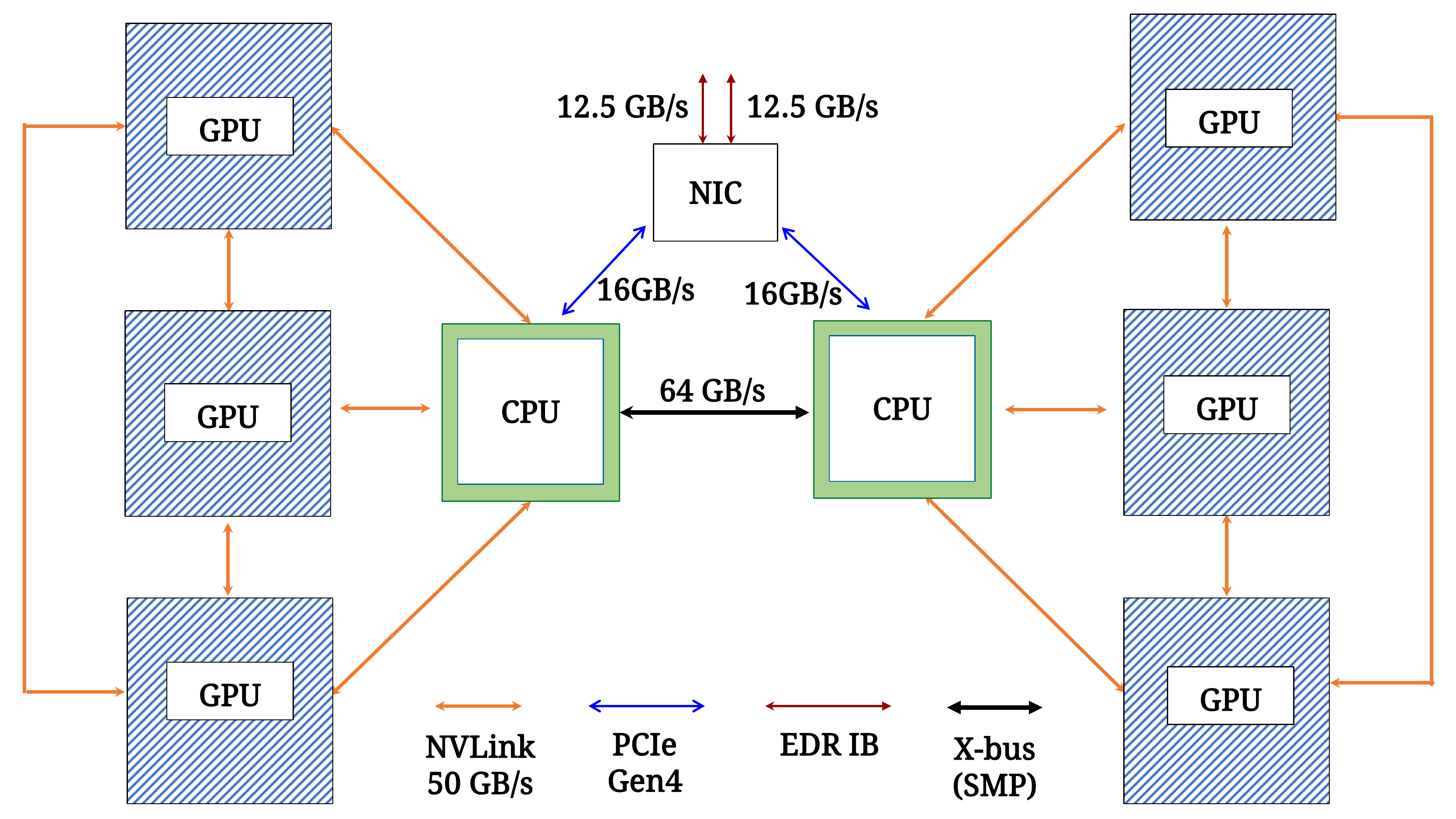}
	\caption{Architectural layout of a single node on Summit.}
	\label{fig:summit_node}
\end{figure}

Summit is a 4,600 node, 200 PFLOPS IBM AC922 
system. 
Each node consists of two \emph{IBM
POWER9} CPUs with 512 GB DDR4 RAM and six NVIDIA V100 GPUs with a total of 96~GB
high-bandwidth memory. 
Each Summit node (Figure~\ref{fig:summit_node}) is divided into two sockets, and each socket has one \emph{IBM
POWER9} CPU and three NVIDIA V100 GPUs, all connected through NVIDIA's high-speed NVLINK2. Each NVLINK2 is capable of a 25 GB/s transfer rate in each direction.
Two \emph{IBM POWER9} CPUs within a Summit node are connected through 
Peripheral Component Interconnect Express bus (64 GB/s bidirectional). 
There is a Mellanox Infiniband EDR network interface connector (NIC) attached to
each Summit node (two ports per NIC, 12.5 GB/s per port).


\subsection{APEX}
APEX~\cite{huck2015autonomic} is a 
performance measurement library for distributed, asynchronous multitasking 
systems. It provides lightweight measurements without perturbing high
concurrency through synchronous and asynchronous interfaces.
To support performance measurement in systems that employ operating system- or user-level 
threading, APEX uses a dependency chain in addition to the call stack to
produce traces and task dependency graphs. 
The synchronous APEX instrumentation application programming interface (API) can be used to add instrumentation to a given
run time and includes support for timers and counters.  To support C++ threads
on Linux systems, the underlying POSIX threads are
automatically instrumented by using a preloaded shared object library that intercepts and wraps
pthread calls in the application. The NVIDIA CUDA Profiling Tools Interface~\cite{cuptiweb} provides CUDA host callback and device activity measurements.
Additionally, the hardware and operating system are monitored through an asynchronous measurement that involves the periodic or on-demand interrogation of the operating system, hardware states, or run time states (e.g., CPU use, 
resident set size, memory ``high water mark''). The NVIDIA Management Library interface~\cite{nvmlweb} provides periodic CUDA device monitoring to APEX.  For this work, APEX was 
extended to capture additional timers and counters related to CUDA device-to-device memory transfers, and support for key MPI calls was provided by
a minimal implementation of the MPI Profiling 
Interface~\cite{MPIbook}.

\subsection{Accuracy Analysis}
To verify that this implementation generates correct results, 
the same input configuration was run  for original and ring algorithm methods, and the differences between the original $G_t$ and accumulated $G^d_t$ arrays were compared. 
A normalized L1 loss function (least absolute deviations, Eq.~[\ref{eq:l1error}]) and normalized L2 loss 
function (least square errors, Eq.~[\ref{eq:l2error}]) were used to compute the normalized error between original $G_t$ 
and accumulated $G^d_t$ arrays in which the ``entrywise''  norm was used.%
\footnote{Entrywise norm as defined in \url{https://en.wikipedia.org/wiki/Matrix_norm}}
The baseline is that the \textit{L1\_error} and \textit{L2\_error} between two arrays 
should be smaller than 5e-7 after DCA++ testing protocol, where:
\begin{equation}
\label{eq:l1error}
    L1\_{\mbox{\textit{error}}} = \frac{\| \mbox{vec}(G_t - G^d_t) \|_{1} }{\| \mbox{vec}(G_t) \|_{1}}
 ~ ,                                      
\end{equation}
\hfill
\begin{equation}
\label{eq:l2error}
    L2\_{\mbox{\textit{error}}} = \frac{ \| \mbox{vec}(G_t - G^d_t) \|_{2} }{ \| \mbox{vec}(G_t) \|_{2} }
~ .
\end{equation}

For input configuration, the single-band Hubbard model was chosen 
because it is a standard model of correlated electron systems 
and is used in almost all the studies of the 
cuprate high-temperature superconductors. 
Moreover, the cluster size was configured to 36-site (6x6 cluster), 
which is state-of-the-art simulations size. 
100,000 Monte Carlo measurements were chosen to observe runtime 
performance of the ring algorithms as the runtime scales linearly 
with the number of measurements for constant number of ranks. 
Since the cluster size was configured to 6*6 and 
four-point-fermionic-frequencies was set to 64, 
this leads 212,336,640 entries in $G_t$. 
Since each $G_t$ entry is a 
double-precision complex number, the $G_t$ memory size is about 3.4 GB. 
This configuration can produce large $G_t$ but still will not hit memory-bound issues 
on Summit GPUs---in which each GPU has 16 GB---for the regular $G_t$ version. 
Such configurations were run on one Summit node five times with six ranks per node and seven walker-accumulator threads per rank. For the distributed $G^d_t$ version, ring size was set to six so there was only one sub-ring during the run. 
The results show that the implementation generates correct results (Table~\ref{table:correctness})
because the \textit{L1\_error} and \textit{L2\_error} on accumulated $G^d_t$ are in an acceptable range.
\begin{table}
\caption{Comparison of function differences between the original $G_t$ and accumulated $G^d_t$ over five runs.}
\label{table:correctness}
\centering
\begin{tabular}{|>{\centering\arraybackslash}m{11mm}
                |>{\raggedleft\arraybackslash}m{30mm}
                |>{\raggedleft\arraybackslash}m{30mm}
                |>{\centering\arraybackslash}m{8mm}|}
\hline
\multicolumn{1}{|>{\centering\arraybackslash}m{11mm}|}{\textbf{Error}} 
    & \multicolumn{1}{>{\centering\arraybackslash}m{30mm}|}{\textbf{Real part}} 
    & \multicolumn{1}{>{\centering\arraybackslash}m{30mm}|}{\textbf{Imaginary part}} 
    & \multicolumn{1}{>{\centering\arraybackslash}m{8mm}|}{\textbf{<5e-7 }}\\
     \hline
     L1 & 3.71e-09$\pm$1.74e-18 & 4.61e-09$\pm$2.16e-18 & True \\
     \hline
     L2 & 3.10e-10$\pm$4.19e-18 & 3.37e-10$\pm$3.89e-18 & True \\
    \hline
\end{tabular}
\end{table}

\subsection{Memory Analysis}
\label{subsec:memory_analysis}
The memory analysis results show that device memory required for $G^d_t$ decreases linearly to the size of the sub-ring or the number of MPI ranks in the sub-communicator, which fits
the ring algorithm.
The APEX profiling tool was used to collect memory allocation information over the time.
The performance results reflect correctly to the ring algorithm method because the $G_t$ was evenly distributed across MPI ranks---in which each rank uses 1 GPU---within one sub-ring communicator. 

\begin{figure}
\centering
     \begin{subfigure}[b]{\columnwidth}
         \centering
         \includegraphics[width=0.4\textwidth, trim=5cm 0 0 0, clip]{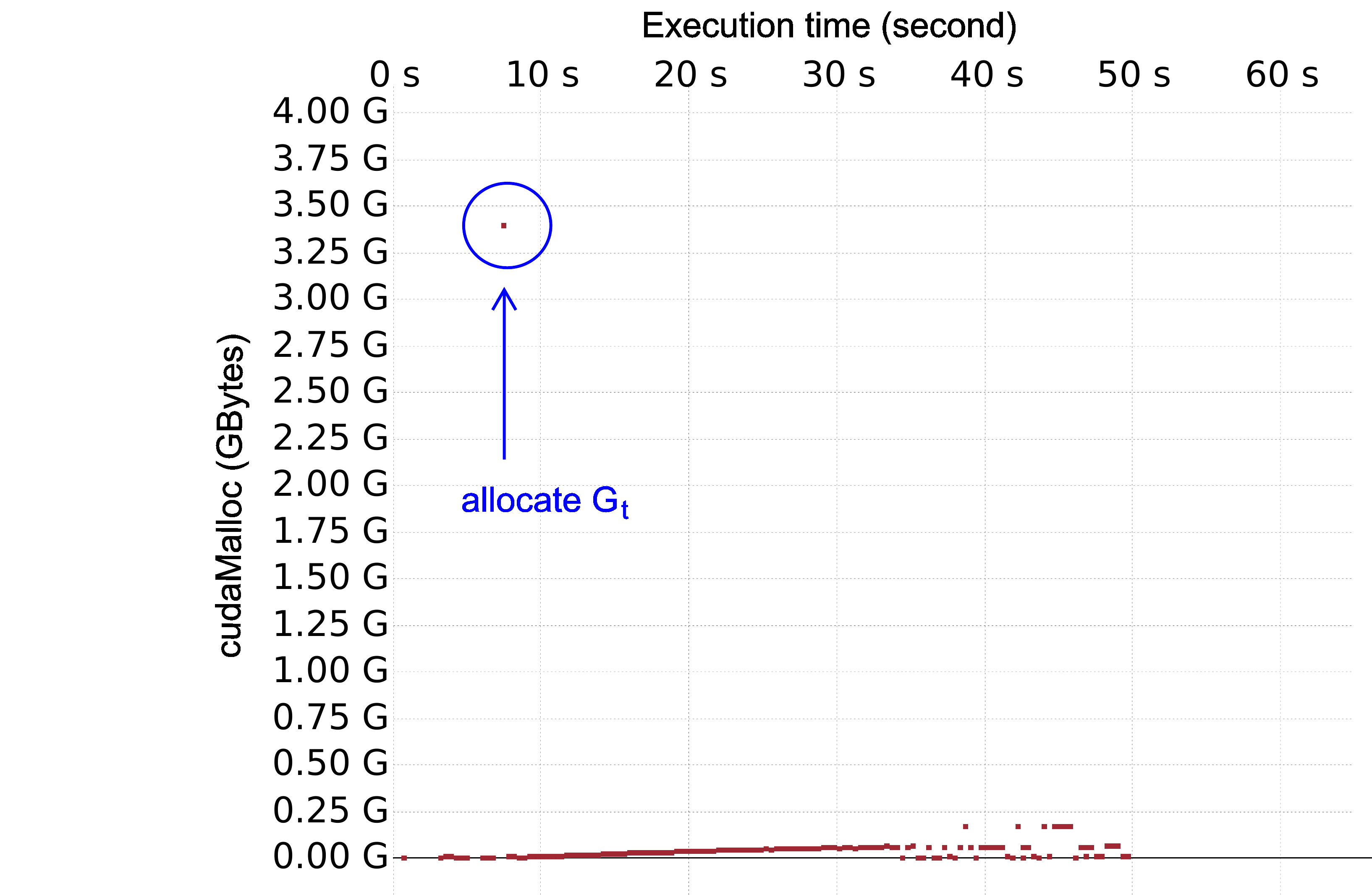}
         \caption{Original $G_t$ implementation.}
         \label{fig:regular_g4}
     \end{subfigure}
     
    \begin{subfigure}[b]{\columnwidth}
         \centering
         \includegraphics[width=0.4\textwidth, trim=5cm 0 0 0, clip]{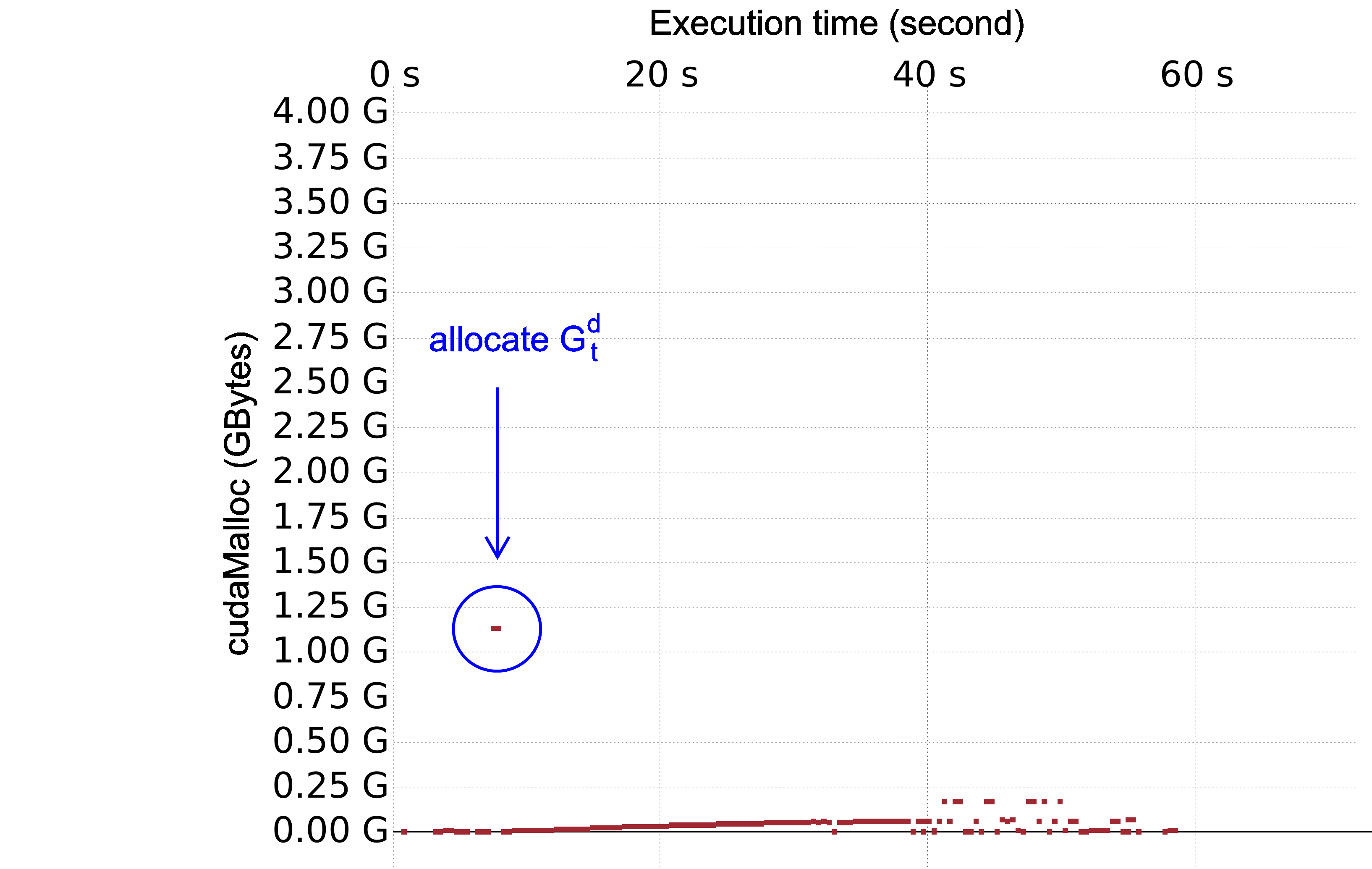}
         \caption{Distributed $G^d_t$ implementation with sub-ring size of three.}
         \label{fig:distributed_g4_bind3GPU}
     \end{subfigure}
     
\caption{cudaMalloc requested size (GB) over time visualized by Vampir.}
\label{fig:cudaMalloc}
\end{figure}

For example, the requested size in \texttt{cudaMalloc} API was compared between original $G_t$ (Figure~\ref{fig:regular_g4}) and distributed $G^d_t$ 
(sub-ring size of three, Figure~\ref{fig:distributed_g4_bind3GPU}) methods. 
This shows that the distributed $G^d_t$ method produced three times less memory allocation 
than the original $G_t$ device array. 
At around 7 s in both cases, the distributed $G^d_t$ method allocated 
1.13 GB for $G^d_{t,i}$, and the original $G_t$ method allocated 
3.40 GB for $G_{t,i}$. 



\subsection{Scaling Results}
\begin{figure}[t]
	\centering
	\includegraphics[width=0.8\columnwidth, trim=2cm 2cm 2cm 2cm, clip]{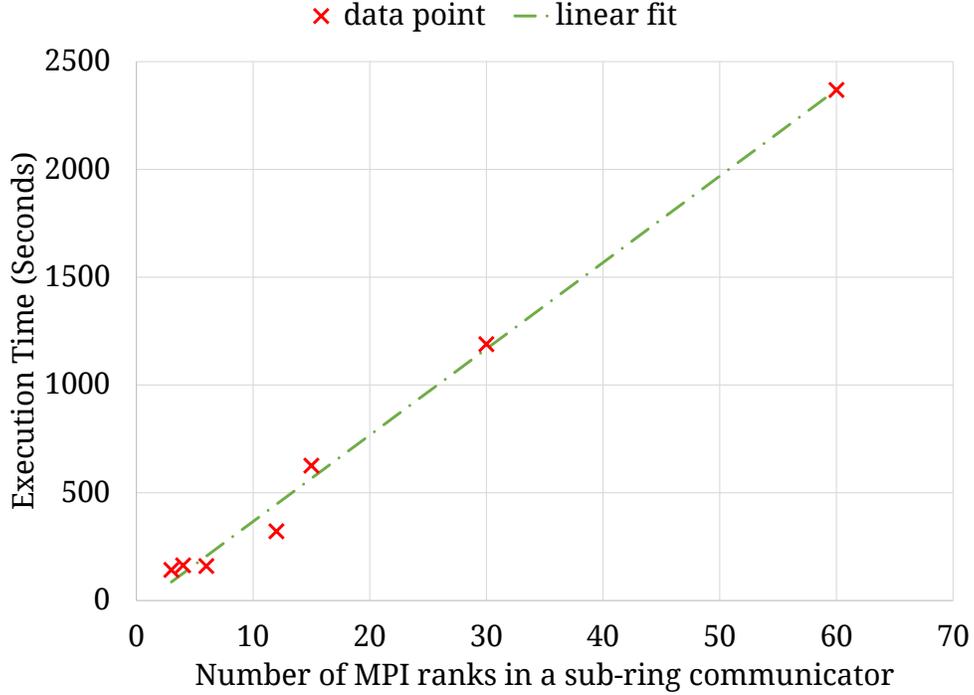}
	\caption{Time for 1,400 iterations (per rank) of the sub-ring algorithm using six ranks per Summit node, and 
	         each message size is 170~MB.}
	\label{fig:subring_scaling}
\end{figure}

%
In the pipeline sub-ring algorithm, each rank sends $S-1$ and receives
$S-1$ messages, where $S$ is the size of sub-ring. Thus, the total
number of messages scales quadratically as O($S^2$), but the number
of messages crossing each communication link increases linearly as
O($S$).
Figure~\ref{fig:subring_scaling} shows the elapsed computation time
for 1,400 measurements (per rank) of the sub-ring algorithm running with six ranks
per Summit node in which each message
is about 170~MB.
The data are well approximated by a linear least-square line that
indicates that the elapsed computation time increases linearly as the sub-ring size increases.
This suggests that the sub-ring algorithm is not constrained by the total
volume of messages but is restricted by the slowest communication link.
The effective bandwidth of the sub-ring algorithm can be estimated as:
\[
\mbox{effective bandwidth} \approx (170*10^6 * S * 1,400)/(\mbox{ elapsed time}),
\]
and this is about 6~GB/s using the data for $S=60$
on 10 nodes
in Figure~\ref{fig:subring_scaling}.
This effective bandwidth is about 50\% of the theoretical peak
bandwidth for the NIC (12.5~GB/s per port)
on the Summit node.

The authors acknowledge that enabling the ring algorithm 
to solve existing small problem-size 
(single band hubbard model with lower cluster size) 
will be an overkill, since the communication overhead 
will drastically increase the runtime; 
therefore the authors propose, the ring algorithm be 
only used when the $G_t$ cannot fit into one single GPU memory. 
When the original DCA++ is executed with a large enough problem size 
(when $G_t$ cannot fit into one single GPU),
the program simply crashes failing to allocate memory on the device. 
Moreover, The scalability issue of the ring algorithm 
was the core focus for the authors during the implementation 
and the optimization design strategies of the sub-ring algorithm. 
Without sub-ring optimization, the originally proposed ring algorithm 
will potentially take undesirably long period of time 
to finish a run of DCA++, 
especially when requesting thousands of compute nodes. 
With the sub-ring optimization, scientists are able to 
run large science cases 
while maintaining acceptable communication overhead. 

Since the current sub-ring size has to be configured manually, 
the authors plan to design a runtime adaptivity optimization 
that will automatically adapt the optimal sub-ring size. 
This optimization will distribute $G_t$
into the minimal number of devices as well as 
preserves optimal runtime performance.
This runtime adaptivity will be very helpful 
because DCA++ is an iterative convergence algorithm 
and thus $G_t$ size could be 
dynamically changed over multiple DCA++ runs 
for production science runs on leadership computing facilities.

%% file: sections/discussion.tex
\section{Discussion}
\label{sec:discussion}

\subsection{Concurrency Overlapping}
The multi-threaded ring implementation provides
sufficient concurrency that overlaps communication and computation.
The APEX profiling tool was used to collect data on process activities over time and visualize the data in Vampir.

\begin{figure*}
	\centering
	\includegraphics[width=\textwidth]{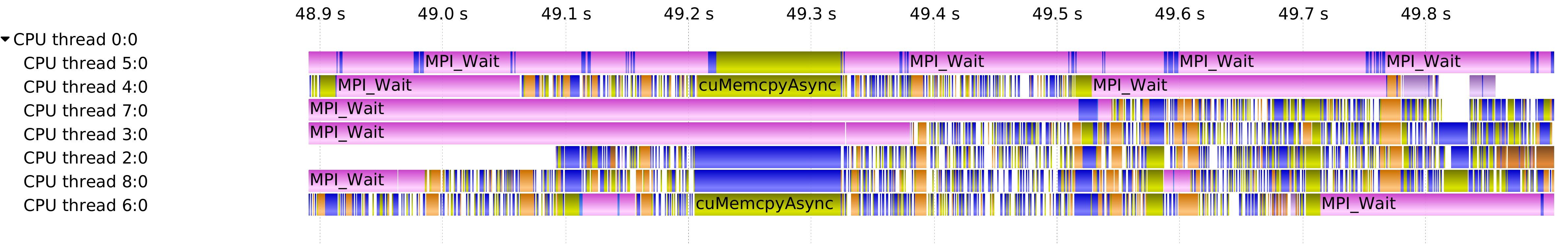}
	\caption{Vampir timeline graph shows the processes' activities over the time in rank 0 (DCA++ with multi-threaded ring algorithm).}
	\label{fig:master_timeline}
\end{figure*}

DCA++ was run with multi-threaded ring support and obtained the timeline activities in rank 0 at 49 s (Figure~\ref{fig:master_timeline}). 
Some concurrency overlap was observed in the multi-threaded
ring algorithm so that although some threads are
blocked in \texttt{MPI\_Wait}, other threads of the same rank perform useful computation tasks.
For example, the short blocks that are not labeled as \texttt{MPI\_Wait} are mostly related to kernel activities.

The current ring algorithm was also observed to be a lock-step algorithm in which the 
next computation (update $G_t$) cannot start until the previous 
communication step ($G_{\sigma}$ message exchange) is finished.
To expose more currency, HPX~\cite{Kaiser2020}---a task-based programming model---could be used to overlap the communication and computation.
For example, DCA++ kernel function can be wrapped into an HPX \textit{future}, which represents
an ongoing computation and asynchronous task. Then, the communication tasks can be attached or chained to the ``futurized'' kernel task. 
Wei et al.~\cite{dca_hpx_2020} reported that DCA++ with HPX user-level~\cite{hpxmp} threading 
support achieves a 20\% speedup over the original C++ threading (kernel-level) due to
faster context switching in HPX threading.

\subsection{Trade-Off between Concurrency and Memory}
As walker-accumulator threads increase in the
multi-threaded ring algorithm,
GPU memory usage is also increased because more device memory is needed to store extra thread-private $G_{\sigma,i}$ buffers. 
This might cause
a new memory-bound challenge if too many concurrent threads are used.
One possible solution is to reduce concurrent threads to achieve more usable device memory.

\begin{figure}
\centering
     \begin{subfigure}[b]{\columnwidth}
         \centering
         \includegraphics[width=0.4\textwidth, trim=5cm 0 0 0, clip]{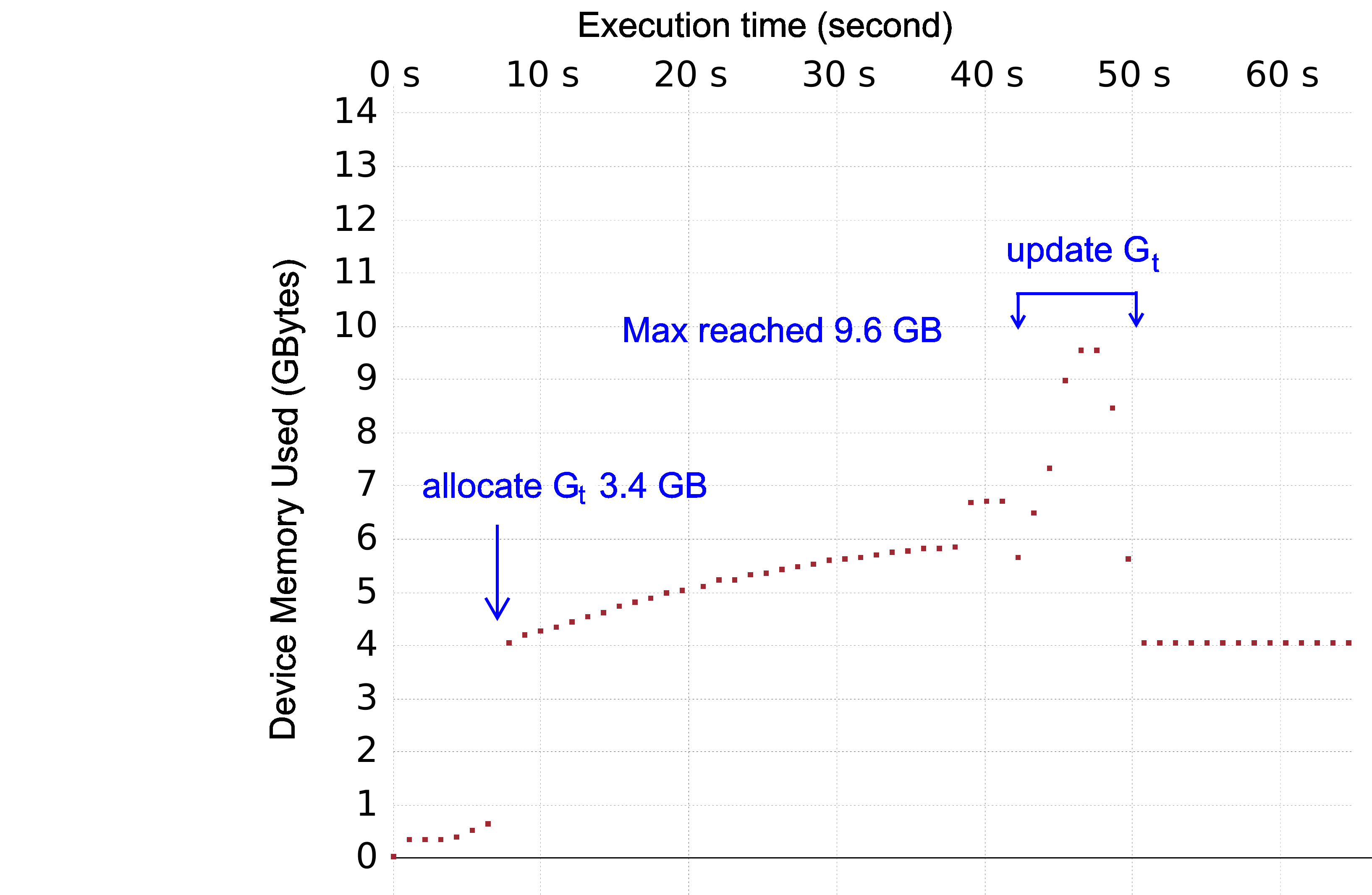}
         \caption{Original $G_t$ method.}
         \label{fig:original_7threads}
     \end{subfigure}
     
    \begin{subfigure}[b]{\columnwidth}
         \centering
         \includegraphics[width=0.4\textwidth, trim=5cm 0 0 0, clip]{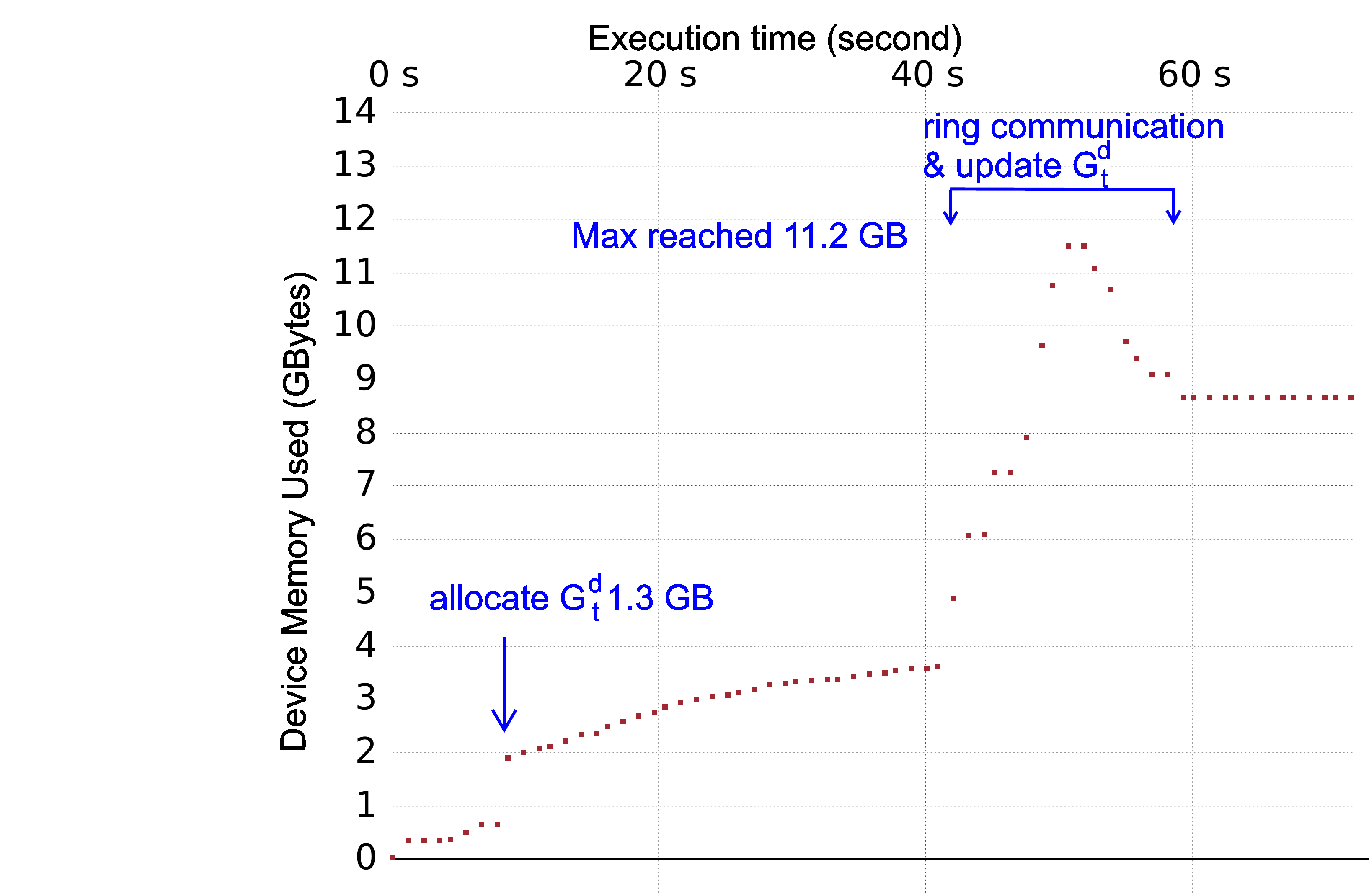}
         \caption{Distributed $G^d_t$ method with sub-ring size of three.}
         \label{fig:distributed_7threads}
     \end{subfigure}
     
\caption{Device memory used (GB over time) when using seven walker-accumulator thread. Visualized by Vampir.}
\label{fig:device_memory_used}
\end{figure}

The same configuration was run for the original $G_t$ and distributed $G^d_t$ versions
with seven threads and then with one thread, respectively (Figure~\ref{fig:device_memory_used}).

For the comparison on seven threads (Figures~\ref{fig:original_7threads} and ~\ref{fig:distributed_7threads}),
the first spike in memory usage increase is due to $G_t$ allocation, and the second significant wave
is because each thread is allocating $G_{\sigma,i}$. 

The original algorithm needs 3.4GB for $G_t$ and 9.6GB in total,
and the new algorithm needs 1.3GB for $G^d_t$ and 11.2 GB in total. 
The non-$G_t$ allocation in the original algorithm is 6.2 GB, 
and distributed $G^d_t$ method is 9.9GB,
which leads to the overhead of 3.7 GB in $G^d_t$ version.
The $G_{\sigma,i}$ is composed of two same-size matrices 
(spin up and spin down matrix, each matrix is sized at 0.17 GB). 
In the original algorithm, the total $G_{\sigma}$ allocation is 
0.17*2*7 = 2.38 GB where 2 is the two matrices (up and down) in 
$G_{\sigma,i}$ and 7 is seven threads. 
In the distributed $G^d_t$ method, 
the total $G_{\sigma}$ allocation is 
0.17*2*3*7 = 7.14GB 
where 3 is three allocations 
($G_{\sigma,i}$ itself, \textit{sendBuf}, \textit{recvBuf}) 
per thread.
The overhead of overall $G_{\sigma}$ allocation 
in the ring algorithm is 7.14 \textminus 2.38 = 4.76 GB, 
which is about 1 GB more than the non-$G_t$ allocation (3.7GB). 
In Figure~{\ref{fig:original_7threads}}, 
there is a significant reduction of allocated memory in the $42^{nd}$ second, which is 1GB memory deallocation in $G_{\sigma}$. 
However, we did not observe a similar drop or wave pattern 
in Figure~{\ref{fig:distributed_7threads}} because those 
\textit{sendBuf}, \textit{recvBuf} matrices are not dynamically 
allocated so that the dip before the allocations was hidden.
This explains the 1GB difference.

\begin{figure}[h]
\centering
     \begin{subfigure}[b]{\columnwidth}
         \centering
         \includegraphics[width=0.4\textwidth, trim=5cm 0 0 0, clip]{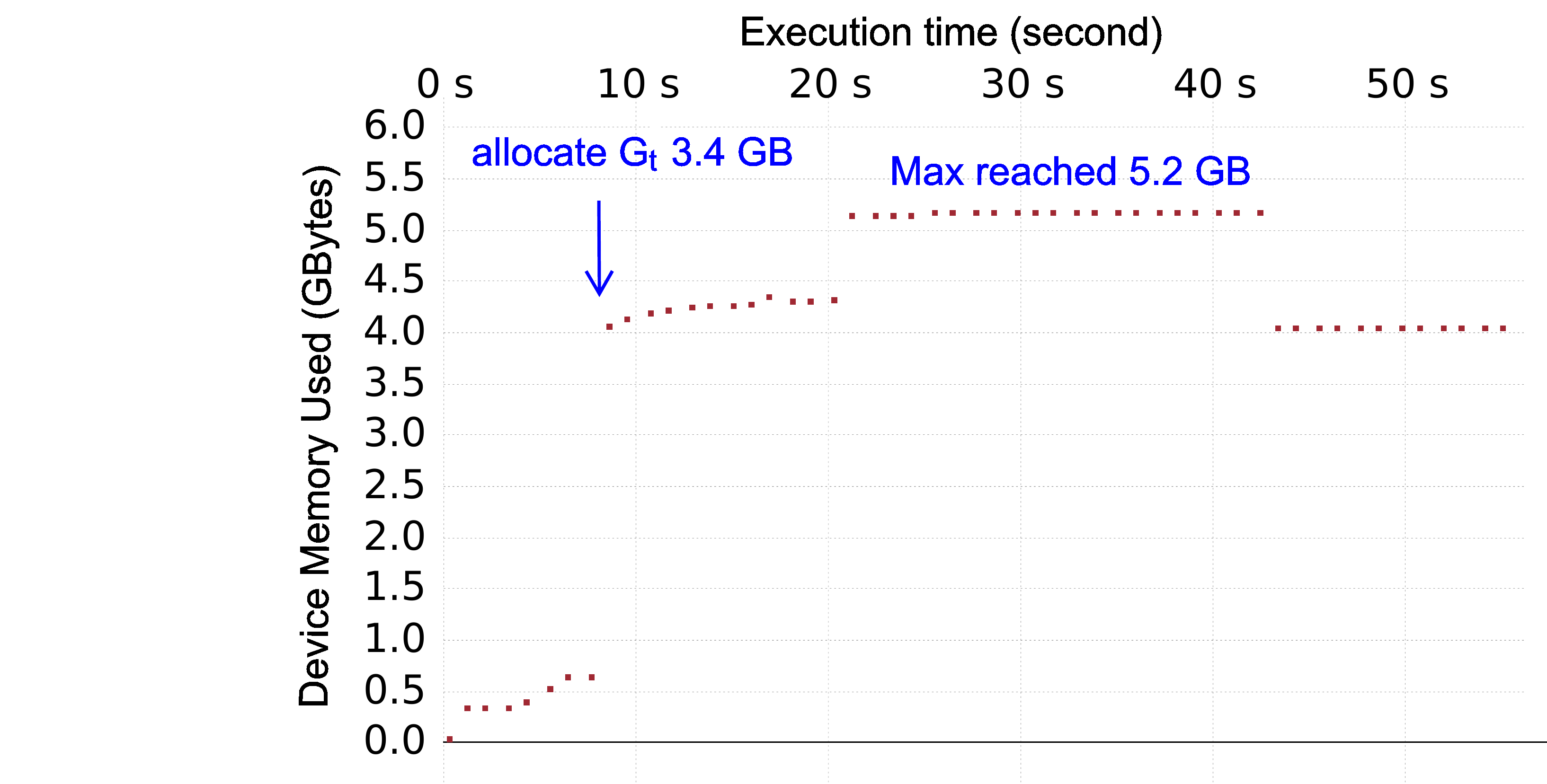}
         \caption{Original $G_t$ method.}
         \label{fig:original_1thread}
     \end{subfigure}
     
    \begin{subfigure}[b]{\columnwidth}
         \centering
         \includegraphics[width=0.4\textwidth, trim=5cm 0 0 0, clip]{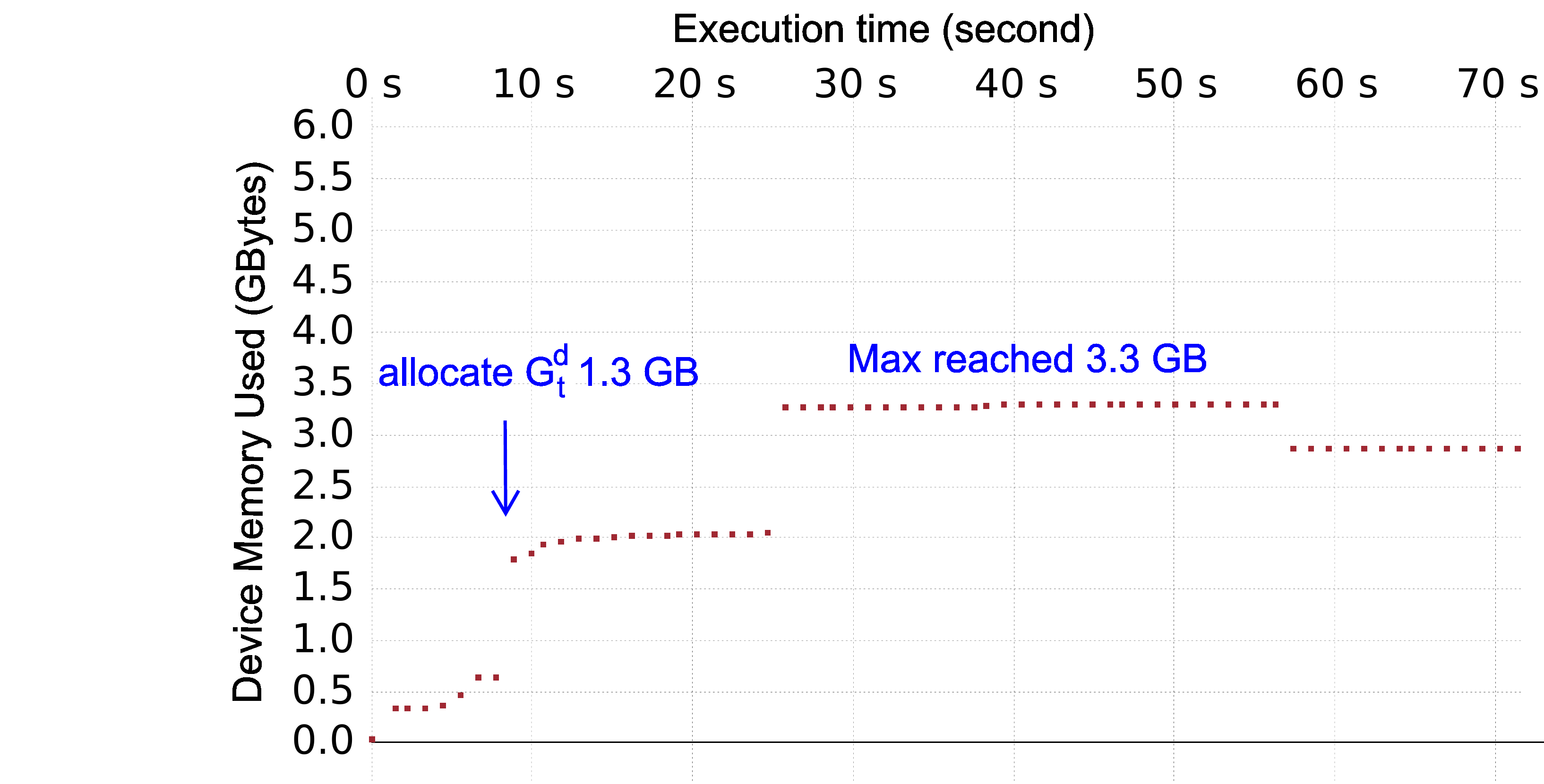}
         \caption{Distributed $G^d_t$ method with sub-ring size of three.}
         \label{fig:distributed_1thread}
     \end{subfigure}
     
\caption{Device memory used (GB over time) when using one walker-accumulator thread. Visualized by Vampir.}
\end{figure}

However, if only one thread was used (Figures~\ref{fig:original_1thread} and ~\ref{fig:distributed_1thread}), then 
the maximum device usage in the distributed $G^d_t$ version (3.3 GB)
is 1.9 GB less than the one in the original $G_t$ version (5.2 GB). Much more usable
device memory can be gained if concurrent walker-accumulator threads are reduced. For example, the saved
device memory from reduced threads can be used to fit larger $G_t$. Furthermore, a comparison experiment was run on one Summit node (six ranks per node) by using the same input configuration 
(sub-ring size is three, measurement is 4,200 total), except for threading numbers per rank. 
The distributed $G^d_t$ with seven threads (87 s) has 1.3 times more speedup 
than the one with one thread (116 s). This result suggests that if there is insufficient 
device memory, then the code might use fewer threads with some loss (less than 30\%) of run time 
performance. The authors are considering quantifying and modeling this trade-off in their future research development.

To solve the NIC bottleneck issue and the new memory-bound challenge
caused by multi-threaded communication (storing additional
$G_{\sigma}$), the authors are considering another plan to move $G_{\sigma}$
to the CPU host in which the CPU host has more memory.
Each Summit node contains 512 GB of DDR4 memory for use by the \textit{IBM POWER9} 
processors, \footnote{Summit User Guide: \url{https://docs.olcf.ornl.gov/systems/summit_user_guide.html} }, 
and there are only 6  * 16 GB = 96 GB of device memory.
On Summit, the NICs are connected to the CPU and are not directly connected to the GPU.
The NVLINK2 connection between CPU and GPU has peak of 50 GB/s, so
it is faster compared with NIC's peak bandwidth (12.5 GB/s) and might not be the bottleneck.
One possible future extension could be to consider keeping $G_t$ on the
CPU side instead of in GPU device memory so that a
smaller sub-ring can be used or so the sub-ring can be kept on the same single node.

Additionally, the authors have explored bidirectional ring implementation 
that alternates ring communication directions between threads. 
After extensive testing, the authors concluded that the 
bidirectional ring improved performance up to 1.3X across-rack 
(each rack has 18 compute nodes on Summit) over the current unidirectional ring.
However, there are no potential performance benefits using the bidirectional ring approach over the current unidirectional ring when reserving the whole rack. 
Authors continue to investigate in coordination with hardware vendors to address the performance of bidirectional ring implementation.

%% file: sections/conclusion.tex
\section{Conclusions}
\label{sec:concl}
This paper presents how the authors successfully solved the
memory-bound challenge in DCA++, which will allow physicists 
to explore significantly large science cases and increase
the accuracy and fidelity of simulations of certain materials.
An effective ``all-to-all'' communication method---a ring abstraction layer---was designed for this purpose so that the distributed device array $G_t$ can be updated across multi-GPUs.
The $G_t$ device array was distributed across 
multi-GPUs so that the allocation size for the most 
memory-intensive data structure per GPU is now reduced to $1/p$ of the original size, where $p$ is number of GPUs in the ring communicator.
This significant memory reduction (much larger $G_t$ capability) 
is the primary contribution 
from this work because condensed matter scientists are now able to 
explore much larger science cases.

In calculating the full 4-point correlation function $G_t$, 
the storage of Gt grows as O( $L^3$ $F^3$) 
where $L$ is the number of cluster sites 
and $F$ is the number of frequencies. 
This new capability will enable large-scale simulations 
such as  36-site (6x6 cluster) with over 64 frequencies to 
(1) obtain more accurate information, and 
(2) enable resolution of  longer wavelength correlations 
that have longer periodicity 
in real space and which cannot be resolved in smaller clusters. 
The system size can grow fairly large and 
depends on how much memory the leadership computing facilities
can provide. 
Relevant science problems that the domain specialists 
would like to study range in the orders of 
10s-of-100s of Gigabits of $G_t$, 
potentially opening up more research 
into how we can use the host memory without losing performance. 

The ring algorithm implementation takes advantage of 
GPUDirect RDMA technology, which can directly and 
efficiently exchange device memory. 
Several optimization techniques were used to improve the ring algorithm performance,
such as sub-ring communicators
and multi-threaded supports. 
These optimizations reduced communication overhead and 
expose more concurrency, respectively.
Performance profiling tools were also improved, such as APEX, which 
now allows more kernel and communication information
to be captured in-depth.
The ring algorithm was demonstrated to effectively reduce the memory
allocation needed for the $G_t$ device array per GPU. 
This paper also discusses various trade-offs between concurrency and memory 
usage for the multi-threaded ring algorithm and the NIC bottleneck issue.
In the future, the authors plan to explore the HPX run time system to overlap 
the computation and communication in DCA++
to expose more concurrency and asynchronicity.

\balance

%% file: sections/acknowledgement.tex
\section*{Acknowledgement}
\label{sec:ack}

Authors would like to thank Thomas Maier (ORNL), Giovanni Balduzzi (ETH Zurich) for their insights during the optimization phase of DCA++. 

This work was supported by the Scientific Discovery through Advanced Computing (SciDAC) program funded by U.S. Department of Energy, Office of Science, Advanced Scientific Computing Research (ASCR) and Basic Energy Sciences (BES) Division of Materials Sciences and Engineering, as well as the RAPIDS SciDAC Institute for Computer Science and Data under subcontract 4000159855 from ORNL. This research used resources of the Oak Ridge Leadership Computing Facility, which is a DOE Office of Science User Facility supported under Contract DE-AC05-00OR22725, and Center for Computation \& Technology at Louisiana State University.